\documentclass[referee]{aa}
\usepackage{graphics,graphicx}
\usepackage{txfonts}
\usepackage{epstopdf}
\usepackage{natbib}

\begin{document}

\title{Evidence for Orbital Motion of Material close to the 
Central Black Hole of Mrk 766}
\subtitle{}

\author{ T.\ J.\ Turner\inst{1 \and 2} \and L.\ Miller\inst{3} \and 
I.\ M.\ George\inst{1 \and 2} \and  
 J.\ N.\ Reeves\inst{2,4} \\
}

\institute{Dept. of Physics, University 
of Maryland Baltimore County, 1000 Hilltop Circle, Baltimore, MD 21250\\ 
\and Code 662,  Exploration of the Universe Division,   
NASA/GSFC, Greenbelt, MD 20771\\ \and Dept. of Physics, University of Oxford, 
Denys Wilkinson Building, Keble Road, Oxford OX1 3RH\\ \and 
Dept. of Physics and Astronomy, Johns Hopkins University, 3400 N 
Charles Street, Baltimore, MD 21218} 

\date{Received /Accepted } 

\abstract{
Time-resolved X-ray spectroscopy has been obtained for the narrow line
Seyfert galaxy Mrk\,766 from {\em XMM-Newton} observations.  We present
analysis in the energy-time plane of EPIC pn data in the
$4-8$\,keV band with energy resolution $R \simeq 50$. 
A component of Fe\,K$\alpha$ emission detected in the
maps shows a variation of photon energy with time that appears both to
be statistically significant and to be consistent with sinusoidal
variation.  We investigate the interpretation that there exists a
component of line emission from matter in a Keplerian orbit around a
supermassive black hole.  The orbit has a period $\sim 165$\,ks and a
line-of-sight velocity $\sim 13,500$\,km\,s$^{-1}$.  This yields a
lower limit for the central mass of M$ > 4.9 \times 10^{5} {\rm
M}_{\odot}$ within a radius of $3.6 \times 10^{13}$\,cm (2.4\,A.U.).
The orbit parameters are consistent with higher black hole masses, but
the lack of any substantial gravitational redshift of the orbit
implies an upper limit to the black hole mass of $4.5 \times
10^{7}$\,M$_{\odot}$.

}

\titlerunning{Orbital Motion in Mrk 766}
\authorrunning{Turner et al.}

\maketitle

\section{Introduction}

AGN are thought to be powered by accretion of material onto a
supermassive black hole.  The accreted gas has angular momentum, and
so forms a flattened, rotating `accretion disk' as it flows toward the
central black hole. Gravitational potential energy lost as the
material approaches the black hole is converted into kinetic energy
and radiation, with the innermost disk producing frequencies up to
the ultraviolet (e.g. \citealt{hm93}).   The observed X-ray
radiation is most likely produced by inverse Compton scattering of
these `seed' photons by relativistic electrons close to the black
hole, possibly in a corona sandwiching the disk \citep{hm93}. While
some of the X-ray radiation is observed directly, a significant
fraction shines back onto the disk surface and produces a so-called
`reflection spectrum', whose main observable feature below 10\,keV is
Fe K$\alpha$ emission (e.g. \citealt{gf91}).

Intrinsically narrow emission lines from an accretion disk would be
affected by a combination of relativistic Doppler and gravitational
redshift effects.  The strong Fe K$\alpha$ line is emitted via
fluorescence or recombination processes between 6.4 -- 7\,keV
(depending on the ionization-state of the gas) and the observation of
a broad and asymmetric profile was predicted for AGN \citep{fab89}
before being established in data from the {\it ASCA}
satellite \citep{tan95}. {\it ASCA} observations of a sample of AGN
found both broad and narrow components of the Fe K$\alpha$ line to be
commonly observed in AGN \citep{n97}.  Contributions from the inner
disk have long been thought to explain the observed broad component of
the line while the narrow line core has been thought to come from material
much further out (\citealt{fab2000} and references therein).

Continuum components absorbed by ionized gas can account for some of
the spectral curvature observed in the Fe K regime. It has thus been
difficult to distinguish between absorbed continua and a
relativistically-broadened line as explanations of some very broad
spectral signatures observed in that energy-band
(e.g. \citealt{reeves04,t04}).  With the realization that
determination of disk parameters may have been compromised by
confusion with absorption, it has become vital to turn to clearer
signatures of the disk.  Fortunately, another diagnostic has recently
been discovered.  Energy-shifted narrow Fe emission lines were first found in
overlapping {\it Chandra} and {\it XMM-Newton} spectra of NGC 3516
\citep{t02}.  The line energies, combined with their rapid flux
variations indicated these could be the long-sought `Doppler horns'
of emission lines from narrow annuli on the accretion disk, perhaps
illuminated by magnetic reconnection events (e.g. \citealt{mf01}) or
by photon bubble instabilities in the inner disk \citep{bty}.  More
examples have since been reported
(e.g. \citealt{g03,y03,trk,delphine,bianchi04}) and the phenomenon
appears to be a general property of Seyfert galaxies.  A rapid (few
thousand second) energy shift detected in Fe emission from the Narrow
Line Seyfert 1 galaxy (NLSy1) Mrk~766 (z=0.0129; \citealt{smithea87})
was of particular interest \citep{trk}, strengthening the hypothesis
that these lines originate very close to the central black hole.
Detailed calculations of the line profiles expected from orbiting
flares on the accretion disk \citep{dov} are consistent with
observations to date, further supporting the hotspot interpretation.

Here we present an application of a relatively new analysis technique
to the {\it XMM-Newton} (hereafter {\em XMM})
data from the narrow line Seyfert 1 galaxy,
Mrk~766. Previous X-ray observations of Mrk~766 have shown X-ray flux
variability by factors of several on timescales down to a few thousand
seconds (e.g. \citealt{p03}).  The X-ray spectrum is complex, strong
features in the soft X-ray regime have been interpreted both in the
context of emission from an ionized accretion disk and absorption due
to ionized gas along the line-of-sight (e.g. \citealt{sako} and
references within).  Fe K$\alpha$ emission has also been detected, but the 
detailed properties of the line have been difficult to determine \citep{p03}.
Our new analysis entails plotting and statistically evaluating the
data as an intensity map in the energy-time plane.  \citet{iwasawa}
have previously analysed energy-time maps to search for time 
variation in emission from NGC~3516 in
the $5.7-6.5$\,keV range, interpreting this as accretion disk emission.
In this paper we search directly for and evaluate the statistical significance of 
any evidence of orbiting matter close to the central black hole in Mrk~766.

\section{Energy-Time Analysis} 

\subsection{The calculation of energy-time fluctuation maps \label{sec:energytime}} 

{\em XMM} spectra from a 2001 observation of Mrk~766 showed line
emission shifting from rest-energy 5.6\,keV to 5.75\,keV within a few
hours, likely to be Fe emission from very close to the black hole
\citep{trk}.  Analysis of the mean spectrum of this AGN has shown a
broad emission component from ionized Fe \citep{p03}, with equivalent
width $\sim 90$\,eV. The breadth of the line indicates an origin
close to the central black hole, supported by \citet{p03} suggestion
of rapid spectral variability in the Fe K emission.

To test the origin of rapid energy shifts in such spectral features we
have created X-ray intensity maps in the energy-time plane.  We have
analysed {\em XMM} observations with the EPIC pn and MOS detectors
made in 2001 May and 2000 May for which data reduction is described by
\citet{trk}.
 
Photons from the source cell were accumulated in pixels in the
energy-time plane.  The pixel distribution was smoothed by the
instrumental resolution in energy, given by a Gaussian of FWHM $\sim
140$\,eV (appropriate for the single plus double events with the latest
calibration), and by a top-hat function of width 10\,ks in time.  Each
time-slice was background-corrected by subtracting a time-dependent
background spectrum measured in an off-source region on the same
detector chip as the source (typically $\sim 5$\% of the source count
rate).  The source continuum was modelled as an absorbed power-law, of
variable amplitude and slope but time-invariant absorption column
density. This continuum was subtracted, leaving positive and negative
residuals that comprise noise plus any emission or absorption
components on top of the continuum.

Noise on this energy-time fluctuation map was calculated by
propagation of Poisson shot noise errors through the smoothing
process.  The noise model comprises of contributions from the variable
subtracted power-law component, the subtracted background component,
the uncertainty in the background component and from the mean excess
line emission above the continuum. Uncertainty in the determination of
the variable power-law component is also present but causes a
time-dependent uncertainty in the zero-level on the map of
fluctuations in the region of the line rather than a random noise
component in each pixel.  At an energy of 6.6\,keV the fractional
uncertainty in the continuum determination is $\pm 0.011$ which leads
to a fractional uncertainty in the zero-level of the signal-to-noise
maps shown below of $\pm 0.13$, small compared with the observed
values in the Fe line.  This zero-point error can cause small apparent
time-variations in flux of an emission feature but cannot cause energy
shifts in detected features.

The ``signal-to-noise'' (S/N) map presented (Fig. 1a) 
is the ratio of the fluctuation
amplitude to the calculated noise.  Formally, the noise 
is not normally distributed, but as the typical number of photons
contributing to the noise in each smoothed pixel is $\sim 170$, the
ratio may be taken as a good indicator of the statistical
significance of individual smoothed fluctuations.

The smoothed energy-time map is not Fourier band-limited, and in order
to preserve the information content of the data, the maps shown for
Mrk~766 are oversampled by a factor 10 on each axis.  On the maps
presented in Fig. 1 there are $\sim$ 13 independent time bins and 28 
independent energy bins, $\sim 370$ independent bins in each map and
thus we expect $\sim 8$ purely statistical positive fluctuations above
the 2$\sigma$ level randomly distributed across the map, significantly
fewer than the actual observed number of such fluctuations.  

We note here that the relatively high oversampling of the sharp-edged
top-hat time-smoothing function does result in visible noise features
at finer time resolution than 10~ks: the advantage of choosing the
top-hat function however is that regions on the maps that are
separated by more than 10~ks are statistically completely independent,
allowing straightforward evaluation of the significance of fluctuations
that are extended in time.  Features from independent time bins that
form part of a long-lived coherent structure have a higher
significance than in an individual bin, and we discuss such features below.

\subsection{Energy-time fluctuations in Mrk\,766}

Before embarking on the statistical analysis of these data, it is instructive
to make a visual inspection.  
Fig\,1b shows the results for the 2001 May EPIC pn observation of
Mrk\,766 as a flux map, plotted as observed energy versus time.  Most
striking is that Mrk\,766 shows emission from the K-shell of Fe that
has the appearance of wandering in energy over a range
$>$1\,keV. There appears to be at least one distinct coherent
structure tracing out a sinusoidal trail in the energy-time domain
(with mean observed energy $\sim 6.6$\,keV and minimum energy occurring
$\sim 60$\,ks into the observation). Fig\, 1b has a 
y-axis range that optimally displays the trail.  Fig\,1a 
shows the corresponding
signal-to-noise (S/N) map 
with a broader range displayed on the y-axis to illustrate the absence of 
trails far away from the Fe K regime.   The highest S/N observed is 5.2.  There is
no evidence for a dominant time-invariant core component of the line and
no component that appears steady in energy.  S/N changes are evident with
time along the trails: some S/N variations are driven by changes in
the background level and/or in the flux of the underlying power-law
(which varies rapidly).   The most persistent feature traces a sinusoid of period $\sim 150$~ks.
While the assertion of a period is a bold one, especially when it is
comparable to the length of the observation, in these data this is
based upon coherent structures composed of many data points giving the
cycle a strong definition. 
These maps delineate more clearly the rapid variations in
flux and energy of components in the Fe K-regime reported previously
by \citet{p03} and \citet{trk}.  We also examined the corresponding
S/N map for MOS 2 (MOS 1 was in timing mode and those data were not
suitable for making an energy-time map). MOS 2 is not as sensitive as
the pn, with only about 30 percent of the count rate.  The
sinusoidal trails are not bright enough to be clearly detected in the
MOS 2 data, but the strongest features in the pn map are confirmed by MOS 2
and therefore cannot be statistical fluctuations in the pn data.
 If we coadd the pn and MOS data the 
highest signal-to-noise ratio observed is 5.9; however, coadding  
these data is not our preferred method for orbit-determination owing to
uncertainty in the relative energy calibration between the two instruments. 

To be sure that the energy calibration is stable we examined the
simultaneous variations in the peak of the instrumental Au-edge (close
to 2\,keV) for the 2001 observation.  We verified that the photon
energy calibration was stable over most of the observation to
0.03\,keV, with the largest variation being 0.06\,keV, and that the
Au-edge does not show the periodic pattern of energy shift evident in
the Fe line.  In any case, residual (uncalibrated) gain fluctuations
could not produce the multiple-peaked structures seen in the map.

There also exists an earlier 80\,ks observation from 2000 May. That
exposure begins with a pn ``closed calibration'', using the on-board
calibration source for the first 13 ks of the observation. Those data
show the Mn-K line to be steady in peak energy. The pn instrument was
then exposed to Mrk~766 for 24 ks, revealing similar characteristics to those
evident in 2001.

For comparison with Mrk~766 we also show the signal-to-noise in the Fe line 
for NGC~5548 in the energy-time plane (Fig\, 2). In contrast to Mrk~766, 
NGC~5548 shows no significant variations in flux or energy of the Fe line. 
This comparison increases our confidence that the pn instrument behaviour 
is well-understood and well-calibrated. 

To test the interpretation of the data further we have generated a
number of simulated datasets, with the same time-varying continuum and
background as the real data and with a superimposed time-invariant
line spectrum given by the mean line spectrum of the real data,
simulated photons being generated by Poisson sampling.  The
simulations are conservative, in the sense that any Doppler broadening
of the line is also included as a non-varying component.  Visual
inspection indicates that although multiple line features can be seen
arising from the Poisson sampling of the broad, multiply-peaked mean
line profile, long-lived trails are not visually apparent.  This subjective
evaluation is put onto a quantitative basis below.

\section{Statistical testing}

\subsection{Statistical methodology}

The first test one could make in attempting to look for signatures of 
orbiting material would be to test whether the data are consistent with 
the null hypothesis, that such signatures are absent.  
We can construct a null hypothesis by
assuming that all the excess Fe emission has a time invariant
spectrum, and then evaluate whether this hypothesis is acceptable by
measuring the $\chi^2$ goodness-of-fit in independent bins in the
energy-time map.  The result for the Mrk~766 data is that the
statistical significance (i.e. the probability of obtaining the observed
result if the null hypothesis is correct) varies from 
$p=0.01$ to essentially unity, depending on the range of energy that
is covered, the bin interval in energy, and the bin interval and
placement in time.  In fact, the value $p=0.01$ is only achieved for
one particular placing of bins.  A more typical value of $p=0.07$ is
obtained if 8 equal-width time bins are used, sampling the energy
range $6 < E/{\rm keV} < 7.2$ with bin width 0.14\,keV.  We can immediately
see that there are two problems with this approach.  First, if we
allow the data sampling to vary until we find the most extreme value
of $\chi^2$, then we don't know the distribution of the test statistic
and hence the test is invalid.  The only way around this is to choose
a data sampling {\em a priori} that would be most sensitive for
distinguishing between models: that sampling would depend on the
velocity and period of the orbiting material, and hence already we
have to introduce the concept of testing between two hypotheses (``no
orbiting matter'' or ``orbiting matter with a particular velocity and
amplitude'').  Second, the $\chi^2$ test above is only searching for evidence of excess
deviation from the model, calculated in each bin individually, above that expected from noise.
There is no notion of testing for the pattern of behaviour
between different bins, or for some correlation in the signal between
bins.  In the search for line-emitting orbiting matter, we do expect a strong correlation
in the signal between adjacent time slices, with a 
sinusoidal variation in line energy with time.
The $\chi^2$ test as so far discussed makes no use of such information,
and as a result the probability that we would wrongly accept the null
hypothesis (a ``Type II'' error: \citealt{kendall}) may be high
(and hence the ``power'' of the test low).

Rather than simply testing the null hypothesis, in this paper we attempt 
to distinguish between two competing hypotheses: either that
the Fe line emission is essentially time invariant, from a region that
is physically distant from the central X-ray source; or that at least
some of the Fe line emission arises from matter orbiting a
supermassive black hole, 
as has been inferred from the time-averaged spectra of active
galaxies as discussed in the introduction.  To do this we proceed as follows.

\subsection{Searching for signatures of orbiting matter in the data}

We start by constructing models of emission that would be expected from
a small line-emitting hotspot in an accretion disk, fit such a model to the time-resolved data, 
and hence investigate whether there is any evidence for the periodic variation
of emission with time that would be expected.  In reality we might expect a
more complex pattern of emission: each emitting region would have a spectrum
of emission rather than a single line; there could be multiple hotspots; and emitting
regions could be substantially extended.  In any of these cases we would expect
there to be a net signal arising from the model that we have chosen to test
and hence the test still has statistical power.  If the emission were
from a uniform disk, however, no such signature would be detectable.

In order to detect such emission, we first construct the ``null hypothesis''
model comprising a variable continuum and background plus a time-invariant
additional component.  The variable continuum is modelled as a power-law of
variable amplitude and slope.  In practice these quantities need to be determined
from data that has been time-averaged to some extent, and here we choose a 
time-averaging of 5000\,s.  The shorter the averaging time, the shorter the
period of orbiting matter that could be searched for.  But 
if the time-averaging were shorter than 5000\,s, the
uncertainty in continuum level would become significant.  This sets a limit on
the range of detectable periods as described below.  As described in
section\,\ref{sec:energytime}, the continuum is
assumed to be absorbed by a time-invariant cold absorber with a column density
obtained from a fit to the entire dataset.  The best fit absorption column
density is $0.9 \times 10^{22}$\,cm$^{-2}$ and has relatively little effect at
the energies of the Fe K$\alpha$ lines considered here.  The additional
time-invariant component (presumably Fe emission) is empirically determined
by taking the mean of the residual spectrum after subtraction of background 
and absorbed power-law continuum.  Hence we make no assumption about the spectrum
of the time-invariant line emission: its spectrum is allowed to be determined
completely by the data.

We then add a model of emission from an orbiting hotspot to the ``null
hypothesis'' model.  A component of narrow line emission whose energy
varies sinusoidally with time is convolved with the instrumental FWHM
of $0.14$\,keV.  The flux from the hotspot is modulated by the
expected Doppler variation in photon arrival rate, although this is a
relatively small effect at the velocities considered here.  Possible
distortions in the radial velocity curve arising from light-bending
are neglected, which is reasonable at the radii of orbits that are
detectable (as discussed below) provided any orbit is not seen very
close to edge-on.  The model is calculated at high time resolution and
the spectrum of emission in each 5000\,s time slice is created from
that.  The time-invariant Fe component is then corrected for the
presence of the orbiting emission.

We may then calculate the improvement in chi-squared, $|\Delta \chi^2|$,
that results when the hotspot is included in the model.  
As the null-hypothesis model assumes that all the excess Fe emission
is time-invariant, this statistic tells us the improvement in the goodness-of-fit
that results from including emission from orbiting material.  
We should note that $|\Delta \chi^2|$ is equal to the square of the signal
to noise of the inverse-variance weighted flux in the orbit.  Because of
the requirement for the time-averaged residual spectrum always to be zero,
all pixels on the energy-time plane, within the energy range covered by the orbit,
are affected by the addition of the model, and $|\Delta \chi^2|$ has to be
calculated over that entire range of energy-time pixels.

The model has four free parameters: the rest-energy of the line $E$; the
line-of-sight velocity amplitude $v_{\rm LOS}$; the angular frequency of the
variation $\omega$; and the phase of the sinusoid $\phi$.  The flux in the hotspot is
assumed unknown, and in calculating $|\Delta \chi^2|$ for any combination of
the first four parameters we adopt the
flux value that maximises $|\Delta \chi^2|$ (for a Gaussian likelihood
function this is equivalent to treating the hotspot flux as a
``nuisance'' parameter and marginalising over it).  Only positive
values of flux are allowed.

In order to evaluate the models we need to place priors on the model
parameters, but the range of model parameters has to be chosen to be
sufficiently broad to encompass the full range in which we would be
interested.  The principal Fe line emission in Mrk~766 occurs over the
energy range $6.4-6.8$\,keV, but in general K$\alpha$ emission from
neutral and ionised Fe can arise over a range of energies from
$6.4-6.96$\,keV \citep{v96}.  In addition, at radii close to a black
hole the line energy is decreased by the gravitational redshift factor
$\sqrt{1-2r_g/r}$, where $r_g = GM/c^2$.  In this paper we investigate
two choices of prior on the line rest-energy.  The broadest choice
allows the line rest-energy in the rest-frame of Mrk\,766 to vary
between 6.07 and 6.96\,keV, thereby covering all possible emission
from radii $r > 20 r_g$ and from ionised Fe at smaller radii also.
The second choice is the narrower range of $6.4-6.8$\,keV which would
be the most sensitive range for investigating the hypothesis that a
significant fraction of the {\em observed} line emission comes from
orbiting material.
  
The velocity amplitude is allowed to vary over the range $0 < \beta \leq 0.15$, where
$\beta \equiv v_{\rm LOS}/c$; and the phase over the range $0-2\pi$.
The necessity to time-average the data sets a limit on the maximum
angular frequency that can be detected.  For an energy resolution
$\Delta E$, time-averaging of $\Delta t$ causes
degradation of the signal when 
$\beta \omega \ga \left(\Delta E/E\right) \left(\Delta t\right)^{-1}$, 
and hence we place an upper limit on $\omega$ of 
$2 \left(\Delta E/E\right)\left(\beta \Delta t\right)^{-1}$.  
We also place a lower limit of $50$\,ks on
the period regardless of velocity. These constraints on $\omega$ and
$\beta$ mean that all possible orbits that are detectable have been
searched if the black hole mass is in the range $4.7 \times 10^{5} <
M_{\rm BH}/{\rm M}_{\odot}\sin^{-3}i < 1.2 \times 10^{7}$,
where $i$ is the orbit inclination to
the line-of-sight,
thereby covering the possible values of black hole mass that have been
suggested in the literature of $6.3 \times 10^5$\,M$_{\odot}$
\citep{botte}, $4.3 \times 10^6$\,M$_{\odot}$ \citep{wl} and 
$10^7$\,M$_{\odot}$ \citep{w02}, at least for high values of $\sin i$.  
The range of velocity amplitudes searched
implies that we have searched for Keplerian orbits at $r > 44 r_g
\sin^2 i$. 
The lower limit on the period implies that in
the low-mass regime we have searched for Keplerian orbits at
$r > 137 (M_{\rm BH}/10^6 {\rm M}_{\odot})^{-2/3} r_g$.  Finally, we note
that there is a minimum detectable energy shift associated with an orbit,
which arises when the maximum energy variation is comparable to the instrumental
FWHM.  Smaller energy shifts are not distinguishable from the null
hypothesis of no variation with time.  

These limits are displayed in Fig.\ref{fig:datafit}, where the
results from the fitting process are shown (adopting the broad prior on the
line rest-energy). In Fig.\ref{fig:datafit}, 
the four-dimensional parameter space is projected onto two parameters,
$\omega$ and $\beta$, and contours of the maximum projected 
$|\Delta \chi^2|$ improvement are shown.  
The largest $|\Delta \chi^2|$ value is 17.68, corresponding to a
weighted signal-to-noise of 4.2, and its location 
is indicated by a large star symbol on this and subsequent figures.  
It has values $\omega = 0.38\pm 0.02\times 10^{-4}$\,s$^{-1}$,
corresponding to a period of 165\,ks, and $\beta = 0.045\pm 0.005$. 
The fitted orbit is shown overlaid on the pn flux map in 
Fig.\ref{fig:overlay}. 
There is some degeneracy between $\omega$ and $\beta$ which arises
because the orbit's period is rather longer than the duration of the
observation.  Further data could help break such degeneracy.

\subsection{Simulation of the data}
The statistical distribution of $|\Delta \chi^2|$ is not known,
however.  Furthermore, the distribution varies with $\beta$, because,
for a given time and energy resolution, there are many more possible
ways of obtaining a positive value of $|\Delta \chi^2|$ at high
$\beta$ than there are at low $\beta$.  In order to assess the
statistical significance of the candidate orbit identified above, we
have created 3000 random simulations of the Mrk\,766 pn observations.
In creating the simulations every attempt has been made to recreate
the same systematic effects present in the data.  Each simulation has
the same varying continuum and background as the actual data, and the
same censoring of bad data.  The source model consists of the ``null
hypothesis'' model described above (absorbed time-variable power-law
continuum, plus background, plus time-invariant Fe excess), and is
Poisson sampled to recreate the shot noise present in the data.

\subsection{Statistical significance of candidate signatures}
These simulations are then treated as if they were real data.  Each
simulation is put through the same background- and
continuum-subtraction and orbit-fitting procedures as the real data,
and the most significant orbit that is found, as measured by the value
of $|\Delta \chi^2|$, is recorded.

The result from this first stage of analysis is that although orbits
that have high signal-to-noise (high values of $|\Delta \chi^2|$) are
detected in the data, if we allow the wide range of parameters
discussed above, then many of the simulations show similarly high
$|\Delta \chi^2|$ improvements.  In the case of the simulations, these
do not arise from genuine orbits: they arise from the joining up of
random shot noise fluctuations, and given that the shot noise in any
narrow time and energy bin is high, and given that a very wide range
of parameters in four-dimensional parameter space have been tested, it
should not be surprising that such ``random-noise orbits'' are found.

However, it should nonetheless be possible to distinguish between such
artefacts and a genuine orbit, because we might expect a genuine orbit
to display emission continuously over a time interval comparable to
the orbital period, whereas a random-noise orbit would be composed of
a series of discontinuous high noise peaks that happen to lie along
some path on the energy-time plane.  We therefore adopt and test two
measures that should be sensitive to this difference between
random-noise and genuine orbits.

The first is based on noting that, in the simulations, no actual
orbits exist, and any that are found arise from linking up of random
high peaks in the shot noise.  So we would expect such random-noise
orbits to be characterised by having a poor (high) value of $\chi_{\rm
orbit}^2$, where this is defined as the absolute value of $\chi^2$,
not the improvement in $\chi^2$, calculated by comparing model and
data along the region defined by the orbit being investigated.  We
expect that a genuine orbit should have a relatively low value of
$\chi_{\rm orbit}^2$ compared with the random-noise orbits.

The second measure is to attempt to assess the continuity of the orbit
- is the orbit comprised of continuous emission, or is it comprised of
a series of one or more short-lived features?  Again, we expect the
random-noise orbits to be discontinuous, as fluctuations between
adjacent bins have no correlation.  To test this, we have defined a
continuity statistic, $t_c$, which is the maximum duration anywhere
along an orbit where the flux remains at greater than half the maximum
value of the orbit.  In practice, the presence of noise will cause
even a real orbit to appear discontinuous unless the data are
smoothed, and results are presented here for smoothing by a top hat in
time, with energy that follows the orbit, of three times the
time-slicing employed above (i.e. a smoothing time of $15$\,ks).  We
could have chosen a number of subtly different ways of defining this
statistic, but the choice made here was defined {\em a priori} before
being tested on the data and was not modified subsequently.
Alternative amounts of time smoothing were tested in order to evaluate
the sensitivity of the statistic to this variable, and little change
was found for time smoothing in the range $10-20$\,ks (for shorter
time-smoothing the data are noisy, for longer smoothing timescales
data and simulations appear smooth).

We now compare results from the simulations with those obtained from
the data.  We select from each simulation the orbit with the highest
value of $|\Delta\chi^2 |$ and measure the other two quantities,
$\chi_{\rm orbit}^2$ and $t_c$.  In evaluating the significance of
these statistics we might be concerned that they could be correlated,
so we first investigate the distribution of $\chi_{\rm orbit}^2$
jointly with $|\Delta\chi^2 |$, as shown in
Fig\,\ref{fig:chisqdist}. For clarity only 1000 simulation points are
shown.  
The candidate orbit from the real data has 
$\chi_{\rm orbit}^2 = 12.36$ and $|\Delta\chi^2| = 17.68$:
these values are also shown as a large star.  In fact, there is
little correlation between $\chi_{\rm orbit}^2$ and $|\Delta\chi^2 |$,
but even so we measure directly the joint probability from these two
statistics by constructing the two-dimensional cumulative probability
distribution and measuring the fraction of points from the simulations
that lie within contours of equal cumulative probability.  Contours
that include $0.9, 0.97, 0.99$ of the simulated data are shown in
Fig\,\ref{fig:chisqdist} (again, results are shown from assuming the
broad prior on rest-energy).
The candidate orbit appears significant at a level $p=0.023$ for the
$6.4-6.8$\,keV range of rest-energy prior, or $p=0.036$ for the
$6.07-6.96$\,keV range.

We now consider the second measure, the orbit duration statistic
$t_c$.  Again, we must check whether this statistic is correlated with
any of the previous ones, and in Fig\,\ref{fig:duration} we show the
orbit duration statistic, $t_c$, plotted against the probability
deduced from the combination of $\chi^2$ values on
Fig\,\ref{fig:chisqdist}.  There appears to be no correlation between
these statistical measures, and hence we may treat the value of $t_c$
as being an independent measure of significance.  In fact, only 
a fraction $0.008$ (narrower range of rest-energy prior) or
$0.005$ (broad range of rest-energy prior) 
of the simulations have $t_c$ values as large as the candidate orbit (which
has $t_c$ almost equal to the maximum possible value set by the
observation duration).  Furthermore, only one of the random-noise orbits
from the simulated data that do have long $t_c$ values has a joint
$\chi^2$ probability value as significant as the candidate orbit.  Hence we
conclude that two independent measures show that an orbit is preferred
in these data compared with the null hypothesis of no time varying Fe
component. The two measures yield independent significance levels of
$p \simeq 0.036$ and $p \simeq 0.005$ (broad prior) or
$p \simeq 0.023$ and $p \simeq 0.008$ (narrower prior) 
and in the remainder of this paper
we shall assume that the orbit does correspond to emission from 
material orbiting the black hole in Mrk~766.  

\section{Discussion}

\subsection{Parameters of the orbiting material}
On the assumption that the orbit found above is indeed 
tracing emission from material
in orbit around the central black hole, we may derive some physical
constraints.  A lower limit to the central mass may be derived if
the orbit's inclination to the line of sight is assumed to be $i=90^{\circ}$, in
which case we find $M > 4.9 \times 10^{5}$\,M$_{\odot}$ if we take the nominal
best-fit values.  At this inclination
the orbital radius, assuming circular orbits, would be 
$3.6 \times 10^{13}$\,cm (2.4\,A.U.).   
If the central mass actually has the value $4.3 \times 10^{6}$\,M$_{\odot}$
\citep{wl} the orbital inclination would be $29^{\circ}$ 
(with an uncertainty dominated
by the value of mass assumed).  In this case the orbital 
radius would be $7.5 \times 10^{13}$\,cm,
5\,A.U.  Quoted in terms of the gravitational radius $r_g$ of the black hole,
the orbital radius is 
$r = {\rm r}_g \beta^{-2} \sin^{2}i = 494 {\rm r}_g \sin^{2}i$, or
$115$\,r$_g$ at $i=29^{\circ}$.

We may also place an upper limit on the black hole mass from the
absence of a substantial gravitational redshift. The best-fit line
rest-energy is $6.59^{+.08}_{-.03}$\,keV in the Earth's frame, or
$6.67$\,keV (coincident with 
the energy of the recombination line from Fe {\sc xxv}), 
corrected to the rest-frame of Mrk~766. If we assume that
the intrinsic line energy is no larger than $6.96$\,keV, an
appropriate upper limit for Fe\,K$\alpha$ emission, this places an
upper limit on the gravitational redshift. and hence a lower limit 
$r \ga 24$\,r$_g$.  Combined with the measured value of 
$\omega$ this places an upper limit on the black hole mass of
$4.5 \times 10^{7}$\,M$_{\odot}$, although it is of course dependent
on the correct measurement of the line rest energy. To obtain a
secure determination of this quantity we need an observation that
spans greater than one orbital time period in order to ensure that
there is no ambiguity in the orbit fit.

\subsection{Evidence for other orbiting material}

Having established that there is evidence for some orbiting material,
we may ask whether there is any evidence for any additional orbits.
There are three other maxima in Fig.\,\ref{fig:datafit} which
have $|\Delta\chi^2|>16$.  Inspection of the corresponding orbits shows
that although these are independent of the principal orbit, they are 
not independent of each other, and correspond to alternative
fits to much the same fluctuations in the energy-time data.  We have 
also measured their $\chi^2_{\rm orbit}$ and $t_c$ values.  The maxima at
$\beta = 0.090, 0.105$ have $t_c = 68435, 65329$s with corresponding significance
levels of 0.13 and 0.15 respectively, and the maximum at $\beta = 0.126$
has $t_c = 30030$s and a significance level of $0.67$.  
Hence there is no evidence from this
analysis that any of these possibilities correspond to genuine orbiting
material.   It is intriguing, however, that the $\beta = 0.105$ and
$0.126$ fits both pass through the short-lived feature noted by
\citet{trk} that is discussed in the next section, and we note that if there
is genuine orbiting material that is highly time-variable, as might be expected
in the inner regions of an accretion disk, then our statistical tests would
be biased against their detection.

It is worth noting that these orbits have very similar best-fit
values of $\omega$ to that of the principal orbit that is present
in the data, although differing $\beta$.  They could not be coplanar
with the principal orbit.  Whether these orbits
are genuine or arise from random noise, it might be unusual to have
such a coincidence in $\omega$.  In fact, the distribution of random-noise orbits
is not uniform over the $\omega-\beta$ plane, and the chance of having
a value of $\omega$ from a random-noise orbit within a range $\pm 0.03
\times 10^{-4}$\,s$^{-1}$ of the principal orbit's value is $7$\,percent, unusual
but not excessively so.

Finally, we have searched for evidence of coplanar orbits.  Again, there
is no significant evidence of multiple coplanar orbits in the data,
although the broad complex pattern of emission that is seen could be
composed of a number of overlapping orbits.  Further data would be required
to test this hypothesis.

\subsection{Emission from material close to the black hole}
There appears to be little evidence for any longer-period components
in Mrk\,766, there may however be shorter-period events with periods
and lifetimes shorter than our minimum detectable timescale.  
There could be a number of reasons
why such inner-disk hotspot emission does not dominate.  First, we
would expect hotspots only to survive at most a few orbital periods
owing to differential rotation, and at small radii the orbital period
is very short.  Second, the necessity to time-smooth the data does
result in a selection effect against detection of short time period
variations as mentioned above.  The high-ionization state of the Fe
emission also suggests that at smaller radii Fe atoms may be fully
ionized such that no line is possible even if the disk extends further
in.  Finally, the existence of a dominant component of emission at
certain radii may indicate the location of a structure in the disk
causing enhanced reflection at that radial location.

But signatures of inner-disk emission may be present in the data
nonetheless.  We note
in particular one event at $t \sim 110$\,ks from the start of the
observation with an observed energy $\sim 5.7$\,keV, previously
identified by \citet{trk}, that appears to survive $< 10$\,ks and that
appears to be rather broad in energy.  This is the signature we might
expect from a short-lived flare from the inner regions of the disk.
Using a more traditional method of time-resolved spectroscopy
\citet{trk} interpreted the data as showing a feature at 5.6\,keV that
moved in energy to 5.75\,keV over a timescale of hours. However, the
more informative method of data analysis employed here indicates that
those two lines may be separate events and not necessarily directly
related.  In any case, more data are required to test the prevalence
of such highly-redshifted lines.  

The similarity of the mean Fe\,K$\alpha$ spectral profiles in Mrk~766 and 
other NLSy1s (e.g. \citealt{t01, com, pat}) suggests the particulars 
of the disk in Mrk~766  may be a characteristic of 
the NLSy1 class and one that could give us insight into 
these systems. For example, a high rate of  
magnetic reconnection events in NLSy1s might naturally 
explain both their rapid X-ray continuum flux 
variability and the nature of the Fe line emission, one might then look for 
a link between high accretion rate and rate of magnetic reconnections. 

\section{Summary and Conclusions}

Energy-time maps from {\it XMM} observations of Mrk~766 reveal 
periodic energy shift in a component of Fe K$\alpha$ emission,
with period $\sim 165$ ks.
This can be interpreted  as evidence for emission from
orbiting gas within $\sim 100$\,r$_g$ of the central black hole. 
It seems likely such material would be part of the accretion disk and 
in the context of that model the data may be explained as a 
hotspot on the disk, illuminated by magnetic reconnection. We have shown that
the strongest energy trail is not likely to be produced as an artefact of
shot noise.  If this interpretation is correct, we obtain 
a lower limit for the central mass M$ \geq 4.9 \times 10^{5}$\,M$_{\odot}$. 
The lack of a large component of gravitational redshift in the 
orbiting emission places an upper limit of M$ \la 4.5 \times 10^{7}$\,M$_{\odot}$
Significant further investigation 
should be possible in the near future using long {\it XMM} exposures
and the high-resolution {\it AstroE2} spectra of Seyfert galaxies.

\section{Acknowledgements}
 
T.J.\ Turner acknowledges support from NASA grant  NNG04GD11G. 
The authors thank the anonymous referee for comments that led to significant  
improvements to the paper.

\clearpage
\newpage

\begin{figure}

\begin{minipage}{8.45cm}{
\resizebox{12cm}{!}{
\includegraphics*{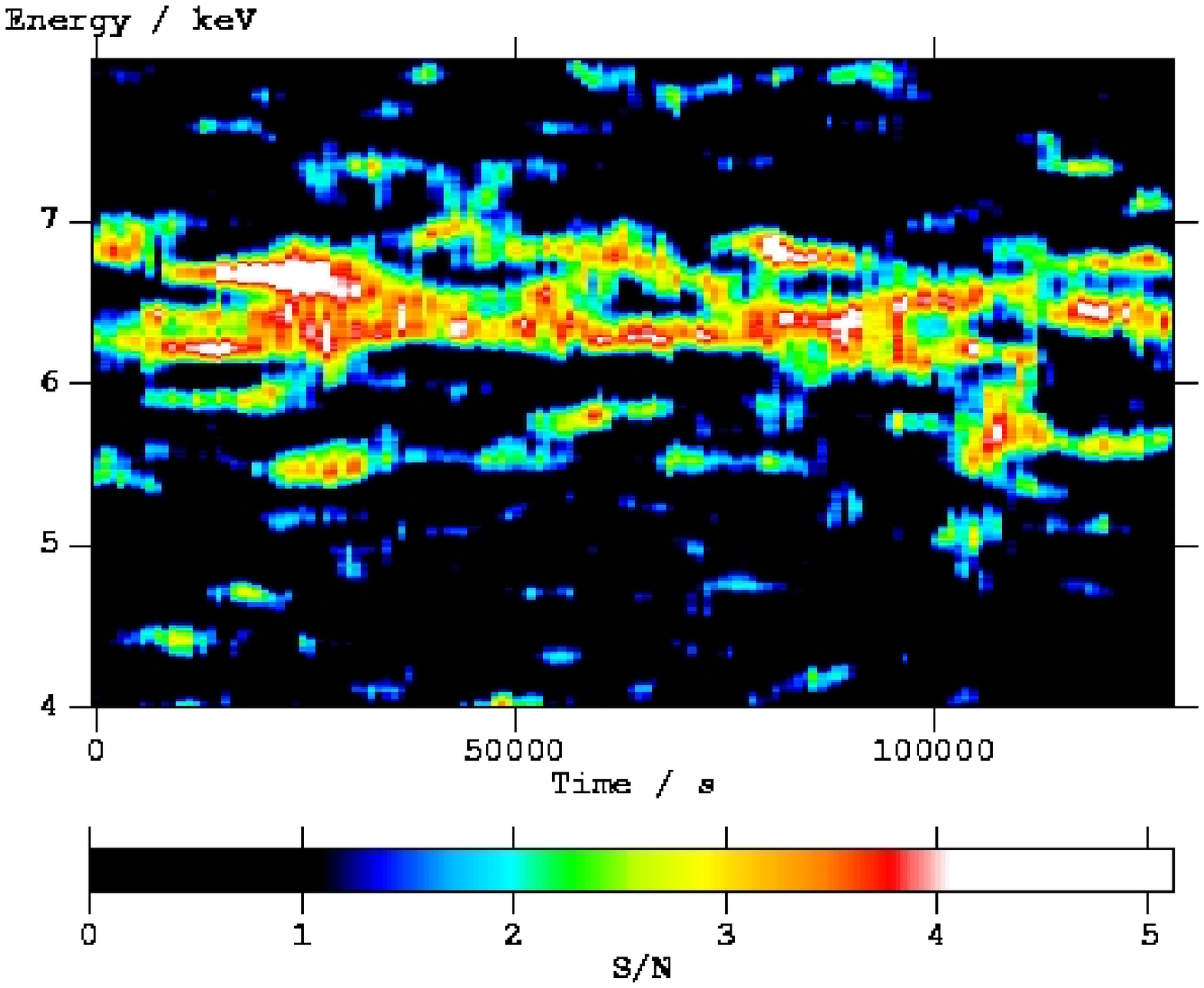}
}
}\end{minipage}

\begin{minipage}{8.45cm}{
\resizebox{12cm}{!}{
\includegraphics*{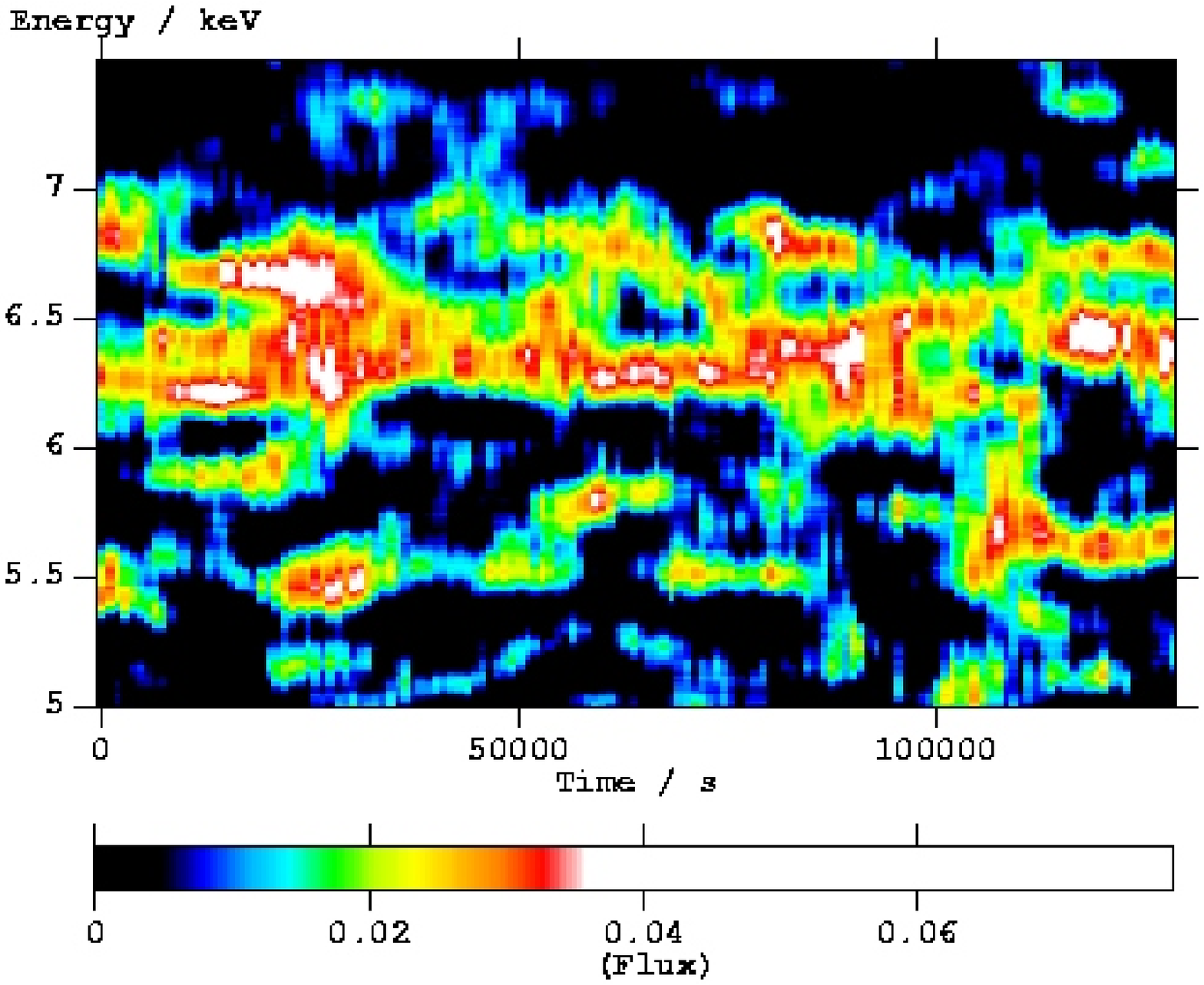}
}
}\end{minipage}

\caption{\it a)  
The signal-to-noise deviations of counts in the Fe-line 
in the energy-time plane for Mrk~766 
 above the power-law continuum.  The
colour-scale represents excess signal in the line counts.  Energy and
time are oversampled by a factor 10.  Data are top-hat smoothed by
10\,ks in time and Gaussian smoothed with FWHM$=0.14$\,keV in energy;
b) the same as a) except these are deviations in the line flux in units of 
count\,s$^{-1}$\,keV$^{-1}$. 
We note the S/N and flux maps have different y-axes, as discussed in \$2.2. 
}
\end{figure}

\begin{figure}
\begin{minipage}{8.45cm}{
\resizebox{12cm}{!}{
\includegraphics[width=8.45cm,angle=0]{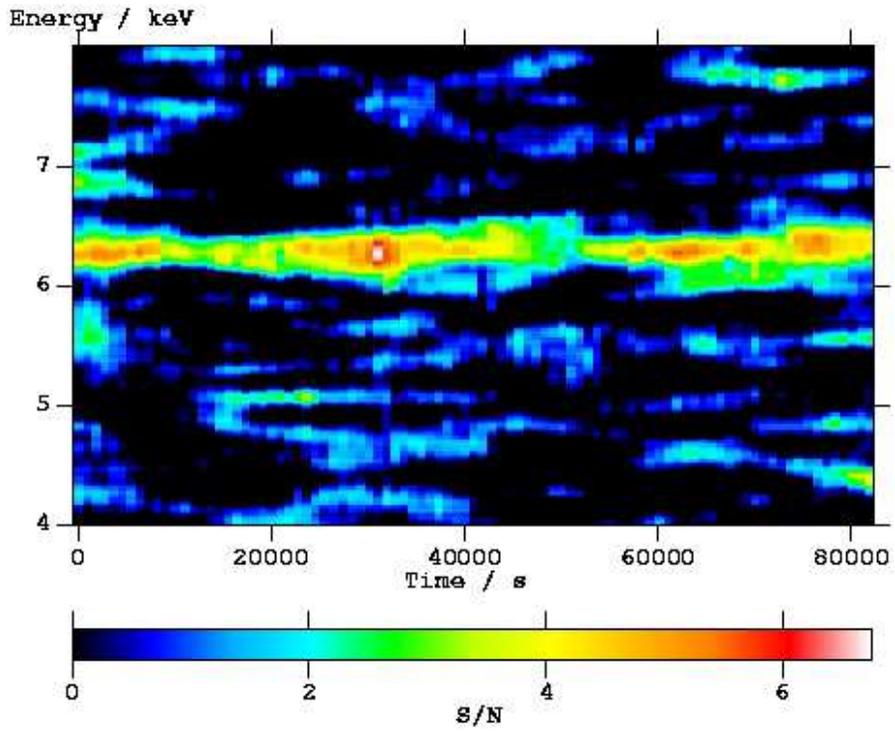}
}
}\end{minipage}
\caption{\it 
Signal-to-noise in Fe line counts above the power-law continuum for NGC 5548. 
}
\end{figure}

\clearpage
\newpage

\begin{figure}
\begin{minipage}{8cm}{
\includegraphics[width=8cm,angle=270]{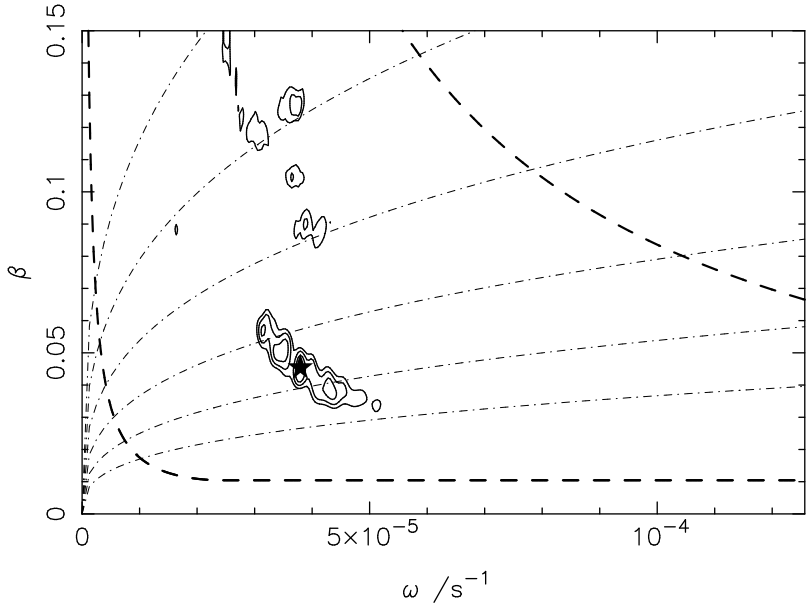}
}\end{minipage}
\caption{\it 
Contours of $|\Delta\chi^2|$ in the $\omega,\beta$ plane.  Contours of
$|\Delta\chi^2| = 15, 16, 17$ are shown.  The maximum value is 17.68.
The most significant orbit is indicated by a large star.  The maximum
values of $\omega$ and $\beta$ that have been searched place limits on
the range of orbital parameters as discussed in the text.  The
right-side dashed line is the locus of maximum detectable $\omega$,
which varies inversely with $\beta$, arising from the time-smoothing
of 5000\,s.  The left-side dashed line is the locus of minimum
detectable energy-shift equal to the instrumental FWHM.  Also shown
are the expected loci of orbits of fixed inclination angle $i$ around
black holes of mass covering the range $10^{5} -
10^{7.5}$\,sin$^{-3}i$\,M$_{\odot}$ (dot-dashed lines, bottom to top
respectively).
\label{fig:datafit}
}
\end{figure}

\begin{figure}
\begin{minipage}{8cm}{
\resizebox{14cm}{!}{
\includegraphics[width=8.45cm,angle=0]{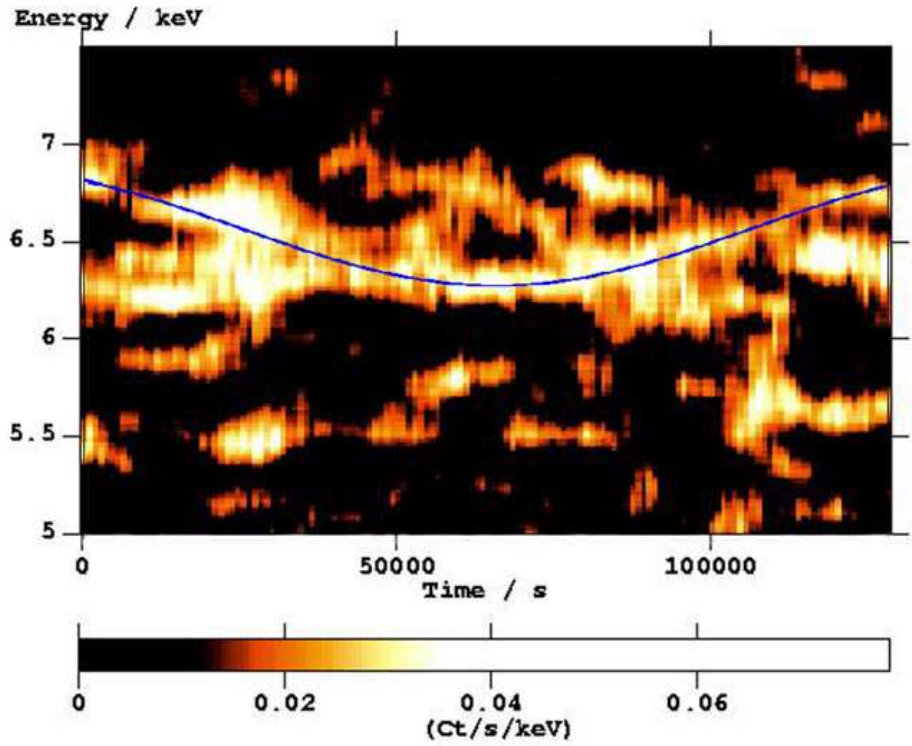}
}
}\end{minipage}
\caption{\it The fitted orbit overlaid on the flux map from the pn data
\label{fig:overlay}
}
\end{figure}

\begin{figure}
\begin{minipage}{8cm}{
\includegraphics[width=8cm,angle=270]{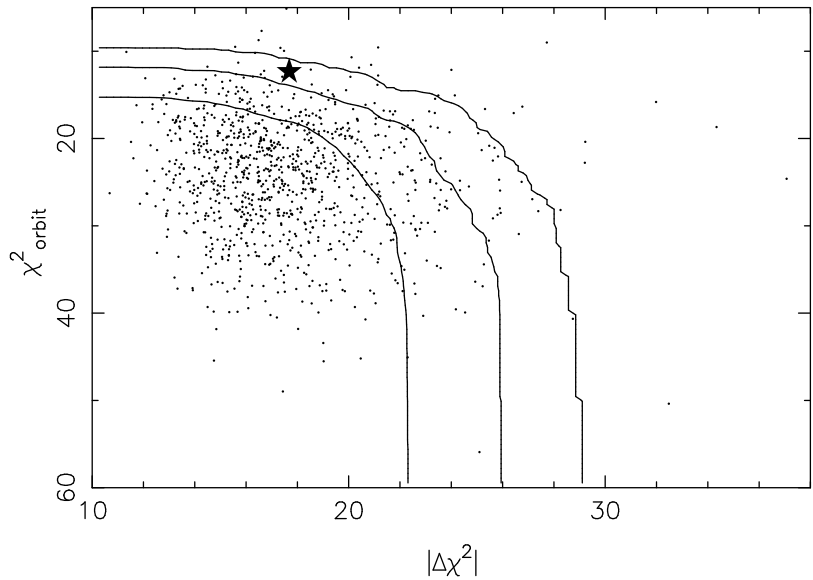}
}\end{minipage}
\caption{\it 
Distribution of $|\Delta\chi^2|$ and $\chi^2_{\rm orbit}$ for the orbit in
each simulation with the largest value of $|\Delta\chi^2|$ (points), assuming
the broad prior on the range of rest-energy (see text).
Contours in this plane that enclose fractions $0.9, 0.97, 0.99$ of the points are
also shown.  Points deviate most from the null hypothesis towards the upper
right of this diagram.  Contours are derived from 3000 simulations: for clarity
points are shown only for 1000 simulations.
The large star denotes the values obtained for the 
orbit associated with the maximum in Fig.\,\ref{fig:datafit}.
\label{fig:chisqdist}
}
\end{figure}

\begin{figure}
\begin{minipage}{8cm}{
\includegraphics[width=8cm,angle=270]{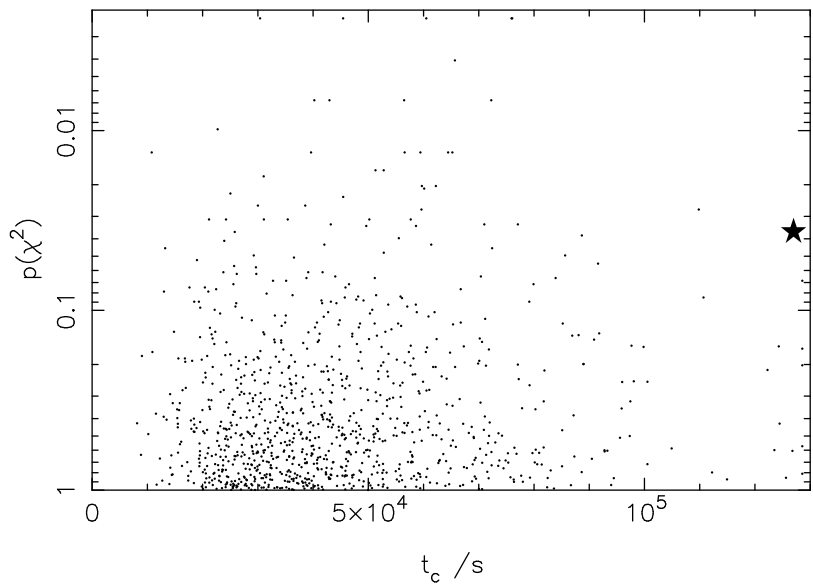}
}\end{minipage}
\caption{\it 
Distribution of probability derived from Fig.\,\ref{fig:chisqdist} plotted
against the trail duration statistic.  Probability values are derived from
3000 simulations: for clarity points are shown only for 1000 simulations.
The large star denotes the values obtained for the 
orbit associated with the maximum in Fig.\,\ref{fig:datafit}.  The broad
prior on the range of rest-energy has been assumed.
\label{fig:duration}
}
\end{figure}

\end{document}